\documentclass[12pt]{amsart}
\usepackage{amsfonts}
\usepackage{layout}
\usepackage[top=2cm, bottom=1.5cm, left=2cm, right=2cm]{geometry}
\usepackage{amssymb}
\usepackage{amsmath}
\usepackage{amsthm}
\usepackage[colorlinks,citecolor=blue]{hyperref}
\usepackage{mathtools}
\usepackage{graphicx}
\usepackage{subfig}
\usepackage{setspace}

\makeatletter

\@addtoreset{equation}{section}
\makeatother

\email{mouayn@usms.ma}
\email{hashim.haydara@gmail.com}
\input{tcilatex}
\begin{document}
\title{Coherent States of Systems with\\ Pure Continuous Energy Spectra}
\author[Z. Mouayn]{Z. Mouayn}
\address{Department of Mathematics\\
Faculty of Sciences and Technics (M'Ghila)\\
Sultan Moulay Slimane University \\
B\'eni Mellal\\
Morocco}
\author[H. A. Yamani]{H. A. Yamani}
\address{Knowledge Economic City, Medina\\
Saudi Arabia}

\date{} 
\maketitle

\begin{abstract}
While dealing with a Hamiltonian with continuous spectrum we use a
tridiagonal method involving orthogonal polynomials to construct a set of
coherent states obeying a Glauber-type condition. We perform a Bayesian
decomposition of the weight function of the orthogonality measure to show
that the obtained coherent states can be recast in the Gazeau-Klauder
approach. The Hamiltonian of the $\ell $-wave free particle is treated as an
example to illustrate the method.
\end{abstract}

\section{\textbf{Introduction}}

Coherent states (CS) have been introduced by Schr\"{o}dinger as states which
behave in many respects like classical states \cite{Sch}. They got this name
after that Glauber \cite{G} realized that they were particularly convenient
to describe optical coherence. In particular, the electromagnetic radiation
generated by a classical current is a multimode coherent state, and so is
the light produced by a laser in certain regimes \cite{Schl}. Therefore, CS
are cornerstones of modern quantum optics \cite{KS} and more recently, CS
found applications in quantum information experiments \cite{GVWB}.

CS also are mathematical tools which provide a close connection between
classical and quantum formalisms so they play a central role in the
semiclassical analysis \cite{AAG, JPG}. \ In general, CS are a specific
overcomplete family of vectors in a Hilbert space associated with a quantum
mechanical system and can be constructed for that space having either a
discrete or continuous basis in different ways \cite{Do} :   ``\textit{\`{a}
la Glauber}" as eigenfunctions of an annihilation operator; as states
minimizing some uncertainty principle or they can be obtained as orbits of a
unitary operator acting on a specific or fiducial state. For the latter one,
Weyl defined them for nilpotent groups \cite{weyl} and this has been
extended to Lie groups \cite{Pe1, O} and further to the continuous spectrum
corresponding to the infinite-dimensional unitary representations of
noncompact groups \cite{Hongoh, Pe2}. 

Unlike the case of systems with pure discrete spectrum, constructing CS for
a pure continuous spectrum is a challenging problem which may be addressed
in different manners but, generally most can be recast in the 
Gazeau-Klauder CS \cite{GK}  which were constructed in terms of the
energy eigenstates of a given non-degenerate system without referring to any
group structure. In \cite{IS} a modification allowing to deal with 
degenerate systems and to treat discrete states and continuous states in a
unified way was proposed.  The problem of building CS from non-normalizable
fiducial states was considered in \cite{BHK}. \ In \cite{PP} the authors
obtained the CS for the continuous spectrum by starting from the
hypergeometric CS for the discrete spectrum, and applying a
discrete-continuous limit. \ In \cite{BK}, the notion of ladder operators was
introduced for systems with continuous spectra together with two kinds of
annihilation operators allowing the definition of CS as modified
eigenvectors of these operators. \ 

Here, our purpose is to construct, under a Glauber-type condition, a set of
CS for a Hamiltonian with continuous spectrum by using the tridiagonal
method. We show that this procedure also tells us how to connect the
constructed CS with their Gazeau-Klauder version for the Hamiltonian under
consideration. This connection is achieved by making appeal to the Bayesian decomposition of
the weight function associated with the  orthogonal polynomials
arising in this method\textbf{.} Indeed, we use the above connection
together with the energy  eigenstates of the non-degenerate system under consideration to show that we recover the
Gazeau-Klauder CS. We illustrate our method for the Hamiltonian
of the $l$-wave free particle.

The paper is organized as follows. In section 2, we introduce a set of
Glauber-type CS by using a tridiagonal method. In Section 3, we recover the
Gazeau-Klauder CS using a Bayesian approach. In Section 4, we illustrate our
method for the Hamiltonian of the $\ell $-wave free particle and we discuss
some of its properties. Section 5 is devoted to some concluding remarks.

\section{\textbf{Glauber-type CS using the tridiagonal approach}}

\subsection{The tridiagonal approach}

Here, we first summarize some needed facts on the tridiagonal method. For
this, we assume that the matrix representation of the given Hamiltonian $H$
in a complete orthonormal basis $\mid \phi _{n}>,n=0,1,2,...$ $,$ is
tridiagonal. That is, 
\begin{equation}
\left\langle \phi _{n}\left\vert H\right\vert \phi _{m}\right\rangle
=b_{n-1}\delta _{n,m+1}+a_{n}\delta _{n,m}+b_{n}\delta _{n,m-1}\text{.} 
\tag{2.1}
\end{equation}%
We now define the forward-shift operator $A$ by its action on the basis $%
\left\vert \phi _{n}\right\rangle $ as follows 
\begin{equation}
A\left\vert \phi _{n}\right\rangle =c_{n}\left\vert \phi _{n}\right\rangle
+d_{n}\left\vert \phi _{n-1}\right\rangle ,\text{ }n=1,2,...\text{ .}   \tag{2.2}
\end{equation}%
For $n=0$, we state that $d_{0}=0.$ Furthermore, we require from the adjoint
operator $A^{\dagger }$ to act on the ket vectors $\mid \phi _{n}>$ in the
following way: 
\begin{equation}
A^{\dagger }\left\vert \phi _{n}\right\rangle =c_{n}\left\vert \phi
_{n}\right\rangle +d_{n+1}\left\vert \phi _{n+1}\right\rangle \text{, \ \ \
\ \ \ \ \ \ }n=0,1,2,3,...\text{ .}  \tag{2.3}
\end{equation}%
The operator $A^{\dagger }A$\ now admits the tridiagonal representation 
\begin{equation}
\langle \phi _{n}\left\vert A^{\dagger }A\right\vert \phi _{m}\rangle
=c_{m}d_{m+1}\delta _{n,m+1}+\left( c_{m}c_{m}+d_{m}d_{m}\right) \delta
_{n,m}+d_{m}c_{m-1}\delta _{n,m-1}  \tag{2.4}
\end{equation}%
in terms of the coefficients $\left( c_{n},d_{n+1}\right) ,n=0,1,2,...$\ .
We have proved \cite{YM} that the coefficients\ in (2.1) are connected to
those in $\left( 2.2\right) $ by the relations $a_{n}=c_{n}c_{n}+d_{n}d_{n}$
and $b_{n}=c_{n}d_{n+1}$, $n=0,1,2,...$ . The tridiagonal matrix
representation of $H$ with respect to the basis $\left\vert \phi
_{n}\right\rangle $ also means that it acts on the elements of this basis as 
\begin{equation}
H\left\vert \phi _{n}\right\rangle =b_{n-1}\left\vert \phi
_{n-1}\right\rangle +a_{n}\left\vert \phi _{n}\right\rangle +b_{n}\left\vert
\phi _{n+1}\right\rangle ,\text{ \ }n=0,1,2,...\text{ .}  \tag{2.5}
\end{equation}%
We may then considered the solutions of the eigenvalue problem $H\left\vert
E\right\rangle =E\left\vert E\right\rangle $ by expanding the eigenvector $%
\left\vert E\right\rangle $ in the basis $\left\vert \phi _{n}\right\rangle $
as 
\begin{equation}
\left\vert E\right\rangle =\sum\limits_{n=0}^{+\infty }\mathcal{C}_{n}\left(
E\right) \left\vert \phi _{n}\right\rangle .  \tag{2.6}
\end{equation}%
Then, making use of $\left( 2.5\right) $, one readily obtains the following
recurrence representation of the expansion coefficients 
\begin{equation}
E\mathcal{C}_{0}\left( E\right) =a_{0}\mathcal{C}_{0}\left( E\right) +b_{0}%
\mathcal{C}_{1}\left( E\right) ,  \tag{2.7}
\end{equation}%
\begin{equation}
E\mathcal{C}_{n}\left( E\right) =b_{n-1}\mathcal{C}_{n-1}\left( E\right)
+a_{n}\mathcal{C}_{n}\left( E\right) +b_{n}\mathcal{C}_{n+1}\left( E\right) ,%
\text{ \ \ \ \ \ }n=1,2,...,  \tag{2.8}
\end{equation}%
and the orthogonality relations 
\begin{equation}
\delta _{n,m}=\int\limits_{\Omega _{c}}\mathcal{C}_{n}\left( E\right) 
\mathcal{C}_{m}\left( E\right) dE,\:n,m=1,2,...\text{ .}  \tag{2.9}
\end{equation}%
These relations correspond to the case when the spectrum of the operator $H
$ is composed only by a continuous part $\Omega _{c}$. Define 
\begin{equation}
P_{n}\left( E\right) :=\frac{\mathcal{C}_{n}\left( E\right) }{\mathcal{C}%
_{0}\left( E\right) },\text{ \ }n=0,1,2,...\text{ .}  \tag{2.10}
\end{equation}%
Then $\left\{ P_{n}\left( E\right) \right\} $ is a set of polynomials that
satisfy the three-term recursion relation for $n\geq 1$ 
\begin{equation}
EP_{n}\left( E\right) =b_{n-1}P_{n-1}\left( E\right) +a_{n}P_{n}\left(
E\right) +b_{n}P_{n+1}\left( E\right)   \tag{2.11}
\end{equation}%
with initial conditions $P_{0}\left( E\right) =1$ and $P_{1}\left( E\right)
=\left( E-a_{0}\right) b_{0}^{-1}.$ If we now define the density $\omega
\left( E\right) :=\left( \mathcal{C}_{0}\left( E\right) \right) ^{2}$ and
assume only existence of continuous spectrum then the relation $\left(
2.9\right) $ reads%
\begin{equation}
\delta _{n,m}=\int\limits_{\Omega _{c}}P_{n}\left( E\right) P_{m}\left(
E\right) \omega \left( E\right) dE.  \tag{2.12}
\end{equation}%
Finally, with the help of the above notations, the coefficients $\left(
c_{n},d_{n}\right) $\ can also be expressed in terms of coefficients $b_{n}$%
\ and the values at zero of consecutive polynomials $\left( P_{n}\right) $
for $n\geq 0$ as\textit{\ } 
\begin{equation}
\left( d_{n+1}\right) ^{2}=-b_{n}\frac{P_{n}\left( 0\right) }{P_{n+1}\left(
0\right) }\text{ \ \ \ \ \ \ \ \ \ }  \tag{2.13}
\end{equation}%
and 
\begin{equation}
\left( c_{n}\right) ^{2}=-b_{n}\frac{P_{n+1}\left( 0\right) }{P_{n}\left(
0\right) }\text{.}  \tag{2.14}
\end{equation}

\subsection{Coherent states}

As in our previous paper \cite{YM} we here adopt the definition of the
Glauber-type CS as the eigenstate of the operator $A$ when the
Hamiltonian is written as $H=A^{\dagger }A.$ \ Note that $A$ is here playing
the role of annihilation operator. Therefore, we first look to the solution
of the eigenproblem 
\begin{equation}
A\left\vert \varphi _{z}\right\rangle =z\left\vert \varphi _{z}\right\rangle 
\tag{2.15}
\end{equation}%
with $z$ real. It is not hard to show that the state satisfying $\left(
2.15\right) $ has the following representation in the chosen basis $%
\left\vert \phi _{n}\right\rangle $ :%
\begin{equation}
\left\vert \varphi _{z}\right\rangle =\left( \mathcal{N}\left( z\right)
\right) ^{-\frac{1}{2}}\sum\limits_{n=0}^{+\infty }Q_{n}\left( z\right)
\left\vert \phi _{n}\right\rangle   \tag{2.16}
\end{equation}%
where 
\begin{equation}
Q_{n}\left( z\right) :=\prod\limits_{j=0}^{n-1}\left( \frac{z-c_{j}}{d_{j+1}}%
\right) ,\text{ \ \ }Q_{0}\left( z\right) =1  \tag{2.17}
\end{equation}%
and assuming that 
\begin{equation}
\mathcal{N}\left( z\right) =\sum\limits_{n=0}^{+\infty }\left\vert
Q_{n}\left( z\right) \right\vert ^{2}<+\infty .  \tag{2.18}
\end{equation}%
\ For $\left( z,\gamma \right) \in \mathbb{R}^{2}$, the generalized CS associated to $H$\ are defined as the orbit of the evolution semigroup $%
e^{-i\gamma H}$\ while acting on the fiducial state $\left\vert \varphi
_{z}\right\rangle .$\ That is, 
\begin{equation}
\left\vert z,\gamma \right\rangle =e^{-i\gamma H}\left\vert \varphi
_{z}\right\rangle .  \tag{2.19}
\end{equation}%
Note that $\gamma $ can be interpreted as a time parameter. \ One observes
from $\left( 2.17\right) $ that when $z=c_{k}$ then the series $\left(
2.16\right) $ terminates and the coherent states reduce to 
\begin{equation}
\left\vert c_{k},\gamma \right\rangle =\left( \mathcal{N}\left( c_{k}\right)
\right) ^{-\frac{1}{2}}\sum\limits_{n=0}^{k}Q_{n}\left( c_{k}\right)
e^{-i\gamma H}\left\vert \phi _{n}\right\rangle .  \tag{2.20}
\end{equation}%
We may also write the $\left\vert \phi _{n}\right\rangle $ as 
\begin{equation}
\left\vert \phi _{n}\right\rangle =\int\limits_{0}^{+\infty }\left\vert
E\right\rangle \sqrt{\omega \left( E\right) }P_{n}\left( E\right) dE 
\tag{2.21}
\end{equation}%
which leads to

\begin{equation}
e^{-i\gamma H}\left\vert \phi _{n}\right\rangle =\int\limits_{0}^{+\infty
}e^{-i\gamma H}\left\vert E\right\rangle \sqrt{\omega \left( E\right) }%
P_{n}\left( E\right) dE.  \tag{2.22}
\end{equation}%
We use the fact that 
\begin{equation}
e^{-i\gamma H}\left\vert E\right\rangle =e^{-i\gamma E}\left\vert
E\right\rangle   \tag{2.23}
\end{equation}%
gives 
\begin{equation}
e^{-i\gamma H}\left\vert \phi _{n}\right\rangle =\int\limits_{0}^{+\infty
}\left\vert E\right\rangle \sqrt{\omega \left( E\right) }P_{n}\left(
E\right) e^{-i\gamma E}dE.  \tag{2.24}
\end{equation}%
Recall that 
\begin{equation}
\langle r\left\vert E\right\rangle =\sqrt{\omega \left( E\right) }%
\sum\limits_{j=0}^{+\infty }P_{j}\left( E\right) \phi _{j}\left( r\right) . 
\tag{2.25}
\end{equation}%
Therefore,%
\begin{equation}
\langle r\left\vert e^{-i\gamma H}\right\vert \phi _{n}\rangle
=\int\limits_{0}^{+\infty }e^{-i\gamma E}\left[ \sum\limits_{j=0}^{+\infty
}\phi _{j}\left( r\right) P_{j}\left( E\right) \right] \omega \left(
E\right) P_{n}\left( E\right) dE  \tag{2.26}
\end{equation}%
\begin{equation}
=\int\limits_{0}^{+\infty }\mathcal{K}\left( r,y\right) P_{n}\left( y\right)
\omega \left( y\right) e^{-i\gamma y}dy  \tag{2.27}
\end{equation}%
where 
\begin{equation}
\mathcal{K}\left( x,y\right) :=\sum\limits_{j=0}^{+\infty }\phi _{j}\left(
x\right) P_{j}\left( y\right) .  \tag{2.28}
\end{equation}%
Finally, summarizing the above calculations, the wave function in Eq.$\left(
2.20\right) $ may also be presented in an integral form as%
\begin{equation}
\langle r\left\vert c_{k},\gamma \right\rangle =\left( \mathcal{N}\left(
c_{k}\right) \right) ^{-\frac{1}{2}}\int\limits_{0}^{+\infty }\mathcal{K}%
\left( r,y\right) \mathcal{S}\left( c_{k},y\right) \omega \left( y\right)
e^{-i\gamma y}dy  \tag{2.29}
\end{equation}%
where 
\begin{equation}
\mathcal{S}\left( u,y\right) :=\sum_{n=0}^{k}Q_{n}\left( u\right)
P_{n}\left( y\right).   \tag{2.30}
\end{equation}

\section{\textbf{Deducing Gazeau-Klauder CS using a Bayesian
analysis}}

\subsection{Bayesian analysis}

Here, our goal is the deduce the Gazeau-Klauder CS \cite{GK}
from the above constructed ones $\left( 2.29\right) $. For that, we assume
that the weight function $\omega _{\lambda }\left( E\right) $ associated
with orthogonal polynomials $\left\{ P_{n}\left( E\right) \right\} $ depends
on a parameter $\lambda $ and that it's a density function for a probability
distribution. Now, the question is to determine two functions: $E\mapsto q\left( E\right) $ and the other $\lambda \mapsto \tau
\left( \lambda \right) $ that may enter in the following decomposition%
\begin{equation}
\frac{\left( \tau \left( \lambda \right) \right) ^{2E}q\left( E\right) }{%
\int \left( \tau \left( \lambda \right) \right) ^{2y}q\left( y\right) dy}%
=\omega _{\lambda }\left( E\right).   \tag{3.1}
\end{equation}%
For that we may look at this problem from a Bayesian viewpoint by saying
that $\left( 3.1\right) $ also means that the weight function 
\begin{equation}
\omega _{\lambda }\left( E\right) \equiv \pi \left( E\rfloor \lambda \right) 
\tag{3.2}
\end{equation}%
can be considered as a \textit{posterior} distribution (or inverse) for an
unknown distribution denoted here by $\pi \left( \eta \rfloor E\right) $
where $E$\ may play the role of a parameter and $\eta $ denotes the variable
or the observed data. We say that $\pi \left( \eta \rfloor E\right) $ is the
statistical \textit{model}. Also from $\left( 3.1\right) $, the unknown
quantity $\left( \tau \left( \lambda \right) \right) ^{2E}$ \ may play the
role of a \textit{prior} distribution on the parameter $E$ which itself is
modeled as a random variable. That is, 
\begin{equation}
\left( \tau \left( \lambda \right) \right) ^{2E}\equiv \pi _{\lambda }\left(
E\right)   \tag{3.3}
\end{equation}%
called\textit{\ }the\textit{\ prior}$.$ According to the general basic
definition (\cite{Rob}, pp.8-10) we also say that $\pi \left( E\rfloor
\lambda \right) $\textit{\ }the \textit{posterior}\ are \textit{conjugate}
under $\pi \left( \eta \rfloor E\right) $\ the\textit{\ model}. Doing so,
our problem in $\left( 3.1\right) $, can be formulated as follows : given $%
\omega _{\lambda }\left( E\right) \equiv \pi \left( E\rfloor \lambda \right) 
$ as posterior, we may ask under which model $\pi \left( \eta \rfloor
E\right) \equiv q\left( E\right) $ the probability law defined by $\omega
_{\lambda }\left( E\right) $ could be conjugate to some prior $\pi _{\lambda
}\left( E\right) $ to be determined ?. \ 

Finally, in concrete situations one will be dealing with the weight function 
$\omega _{\lambda }\left( E\right) $  will be given explicitly
therefore we can find the two quantities $q\left( E\right) $ and $\tau
\left( \lambda \right) $.\ The latter ones, can be used to prove that the
constructed CS we have introduced via the tridiagonal
method procedure agree with the Gazeau-Klauder CS for the
continuous spectrum. Indeed, as we will see below this
analysis will provide us with the factorial function $f\left( E\right) $ and
the re-parametrization formula $s=\tau \left( \lambda \right) $ that serve
as a bridge linking the two approaches.

\subsection{Gazeau-Klauder CS}

Let \ $H>0$ be a Hamiltonian operator with non-degenerate continuous
spectrum and let $\left\{ \left\vert E\right\rangle \right\} $ stands for a
basis of eigenstates in some Hilbert space $\mathcal{H}$, for which 
\begin{equation}
H\left\vert E\right\rangle =E\left\vert E\right\rangle ,\text{ \ \ \ }0<E<%
\overline{E}  \tag{3.4}
\end{equation}%
so that the energy support is $\left[ 0,\overline{E}\right) .$ Here $%
\overline{E}=+\infty $ could be considered. We also can choose a normalized
basis of eigenvectors of $\mathcal{H}$ : 
\begin{equation}
\langle E\left\vert E^{\prime }\right\rangle =\delta \left( E-E^{\prime
}\right)   \tag{3.5}
\end{equation}%
and 
\begin{equation}
\int\limits_{0}^{\overline{E}}\left\vert E\right\rangle \left\langle
E\right\vert dE=\mathbf{1}_{\mathcal{H}}.  \tag{3.6}
\end{equation}%
For $s\geq 0$ and $\gamma \in \mathbb{R}$, the Gazeau-Klauder CS \cite{GK} are defined by 
\begin{equation}
\left\vert s,\gamma \right\rangle =\left( \mathcal{N}\left( s\right) \right)
^{-\frac{1}{2}}\int\limits_{0}^{\overline{E}}\frac{s^{E}}{\sqrt{f\left(
E\right) }}e^{-i\gamma E}\left\vert E\right\rangle dE.  \tag{3.7}
\end{equation}%
These states are normalized 
\begin{equation}
\langle s,\gamma \left\vert s,\gamma \right\rangle =1  \tag{3.8}
\end{equation}%
\ and 
\begin{equation}
\mathcal{N}\left( s\right) =\int_{0}^{\overline{E}}\frac{s^{2E}}{f\left(
E\right) }dE  \tag{3.9}
\end{equation}%
is a normalization factor. The function $E\mapsto f(E)$ is determined by a
suitable non-negative weight function $\sigma (s)\geq 0$ as 
\begin{equation}
f(E)=\int\limits_{0}^{\overline{E}}s^{2E}\sigma (s)ds.  \tag{3.10}
\end{equation}%
With the measure 
\begin{equation}
d\mu (s,\gamma )=\frac{1}{2\pi }\mathcal{N}(s)\sigma (s)dsd\gamma , 
\tag{3.11}
\end{equation}%
the resolution of the identity reads 
\begin{equation}
\int \left\vert s,\gamma \right\rangle \left\langle s,\gamma \right\vert
d\mu \left( s,\gamma \right) =\mathbf{1}_{\mathcal{H}}.  \tag{3.12}
\end{equation}%
Finally, from the above Bayesian decomposition of $\omega _{\lambda }\left(
E\right) ,$ we choose $f(E)$ to be the inverse of $q\left( E\right) $, i.e.,%
\begin{equation}
f\left( E\right) \equiv \frac{1}{q\left( E\right) }  \tag{3.13}
\end{equation}%
and we take 
\begin{equation}
s\equiv \tau \left( \lambda \right)   \tag{3.14}
\end{equation}%
as a new parametrization.

\section{\textbf{Coherent states associated with the $\ell $-wave free
particle\textbf{\ }}}

We start with separating the angular part of the wavefunction of the free
particle in terms of the spherical harmonics that are eigenfunctions of the
angular momentum which is conserved for this kind of potentials. That leaves
for the radial part of the wavefunction the Schr\"{o}dinger operator 
\begin{equation}
H_{\ell }:=-\frac{1}{2}\frac{d^{2}}{dr^{2}}+\frac{1}{2}\frac{\ell \left(
\ell +1\right) }{r^{2}}  \tag{4.1}
\end{equation}%
which acts on the Hilbert space $\mathcal{H}:=L^{2}\left( \mathbb{R}%
_{+},dr\right) $ and admits a continuous spectrum $E\in \left[ 0,+\infty
\right) $. Hence it is positive semi-definite. Here, the oscillator space $%
\mathcal{H}$ is endowed with the orthonormal basis whose elements are given
by 
\begin{equation}
\phi _{n}^{(\ell ,\lambda )}\left( r\right) :=\sqrt{\frac{2\lambda n!}{%
\Gamma \left( n+\ell +\frac{3}{2}\right) }}\left( \lambda r\right) ^{\ell
+1}\exp \left( -\frac{\lambda ^{2}}{2}r^{2}\right) L_{n}^{\left( \ell +\frac{%
1}{2}\right) }\left( \lambda ^{2}r^{2}\right) ,\text{ \ }n=0,1,2,..., 
\tag{4.2}
\end{equation}%
$r\in \mathbb{R}_{+}$ where $\lambda $ denotes a real free parameter, $\ell $
is the angular momentum number and $L_{n}^{\left( \sigma \right) }\left(
.\right) $ is the Laguerre polynomial (\cite{GR}, p.1000). Using
differential recurrence relations for these polynomials, one finds by direct
calculations that the matrix elements defined by 
\begin{equation}
\langle \phi _{n}^{(\ell ,\lambda )}\left\vert H_{\ell }\right\vert \phi
_{m}^{(\ell ,\lambda )}\rangle =\int\limits_{0}^{+\infty }\phi _{n}^{(\ell
,\lambda )}\left( r\right) H\left[ \phi _{m}^{(\ell ,\lambda )}\right]
\left( r\right) dr  
\end{equation}%
have the following expression%
\begin{equation*}
\langle \phi _{n}^{(\ell ,\lambda )}\left\vert H_{\ell }\right\vert \phi
_{m}^{(\ell ,\lambda )}\rangle =\frac{\lambda ^{2}}{2}\left( 2n+\ell +\frac{3%
}{2}\right) \delta _{n,m}+\frac{\lambda ^{2}}{2}\sqrt{n\left( n+\ell +\frac{3%
}{2}\right) }\delta _{n,m+1}\tag{4.3}
\end{equation*}%
\begin{equation*}
+\frac{\lambda ^{2}}{2}\sqrt{\left( n+1\right) \left( n+\ell +\frac{3}{2}%
\right) }\delta _{n,m-1}.
\end{equation*}%
Therefore, we can identify the coefficients $\left( a_{n}\right) $ and $%
\left( b_{n}\right) $ in $\left( 2.1\right) $ as follows 
\begin{equation*}
a_{n}=\frac{\lambda ^{2}}{2}\left( 2n+\ell +\frac{3}{2}\right)\tag{4.4a} ,
\end{equation*}%
\begin{equation*}
b_{n}=\frac{\lambda ^{2}}{2}\sqrt{\left( n+1\right) \left( n+\ell +\frac{3}{2%
}\right) }.\tag{4.4b}
\end{equation*}%
According to equation $\left( 4.3\right) $, the recursion relation is
solved, 
\begin{equation}
P_{n}\left( E\right) =\left( -1\right) ^{n}\sqrt{\frac{n!\Gamma \left( \ell +%
\frac{3}{2}\right) }{\Gamma \left( n+\ell +\frac{3}{2}\right) }}%
L_{n}^{\left( \ell +\frac{1}{2}\right) }\left( \frac{2}{\lambda ^{2}}%
E\right) . \tag{4.5} 
\end{equation}%
These polynomials satisfy the orthogonality relations 
\begin{equation}
\int\limits_{0}^{+\infty }P_{j}\left( E\right) P_{k}\left( E\right) \omega
_{\ell ,\lambda }\left( E\right) dE=\delta _{j,k}  \tag{4.6}
\end{equation}%
with respect to the weight function 
\begin{equation}
\omega _{\ell ,\lambda }\left( E\right) =\frac{2}{\lambda ^{2}\Gamma \left(
\ell +\frac{3}{2}\right) }\left( \frac{2}{\lambda ^{2}}E\right) ^{\ell +%
\frac{1}{2}}\exp \left( -\frac{2}{\lambda ^{2}}E\right) .  \tag{4.7}
\end{equation}%
Note that $\omega _{\ell ,\lambda }\left( E\right) $\ is the continuous
density function of the Gamma probability distribution $G\left( \alpha
,\beta \right) $ with the shape parameter $\alpha =\ell +\frac{3}{2}$ and
the scale parameter $\beta =2\lambda ^{-2}.$ Finally, with%
\begin{equation}
P_{n}\left( 0\right) =\left( -1\right) ^{n}\sqrt{\frac{\Gamma \left( n+\ell +%
\frac{3}{2}\right) }{n!\Gamma \left( \ell +\frac{3}{2}\right) }}  \tag{4.8}
\end{equation}%
equations $\left( 2.13\right),\left( 2.14\right) $ together with $\left(
3.20\right) $  yield 
\begin{equation}
d_{n+1}=\frac{\lambda }{\sqrt{2}}\sqrt{n+1},\text{ \ }c_{n}=\frac{\lambda }{%
\sqrt{2}}\sqrt{n+\ell +\frac{3}{2}}.  \tag{4.9}
\end{equation}%
The kernel function $\mathcal{K}$ in $\left( 2.28\right) $ has the form%
\begin{equation}
\mathcal{K}\left( r,E\right) =\sum\limits_{j=0}^{+\infty }\phi _{j}^{(\ell
,\lambda )}\left( r\right) P_{j}\left( E\right)   \tag{4.10}
\end{equation}%
\begin{equation}
=\left( \lambda r\right) ^{\ell +1}e^{-\frac{1}{2}\left( \lambda r\right)
^{2}}\sqrt{2\lambda \Gamma \left( \ell +\frac{1}{2}\right) }%
\sum\limits_{j=0}^{+\infty }\frac{j!\left( -1\right) ^{j}}{\Gamma \left(
j+\ell +\frac{3}{2}\right) }L_{j}^{\left( \ell +\frac{1}{2}\right) }\left(
\left( \lambda r\right) ^{2}\right) L_{j}^{\left( \ell +\frac{1}{2}\right)
}\left( \frac{2E}{\lambda ^{2}}\right) .  
\end{equation}%
We now make use of the formula%
\begin{equation}
\sum\limits_{j=0}^{+\infty }\frac{j!\left( -h\right) ^{j}}{\Gamma \left(
j+\nu +1\right) }\left( -h\right) ^{j}L_{j}^{\left( \nu \right) }\left(
x\right) L_{j}^{\left( \nu \right) }\left( y\right) =\frac{e^{\left(
x+y\right) \frac{h}{1+h}}}{1+h}\left( xyh\right) ^{-\frac{\nu }{2}}J_{\nu
}\left( 2\sqrt{xy}\frac{h^{\frac{1}{2}}}{1+h}\right)   \tag{4.11}
\end{equation}%
(see \cite{Buch}, p.139, (12a) and p.140 for $h\rightarrow 1)$ for $\nu
=\ell +\frac{1}{2}$, $x=\lambda ^{2}r^{2}$ and $y=2\lambda ^{-2}E,$ we get
that%
\begin{equation}
\mathcal{K}\left( r,E\right) =\frac{1}{2}\lambda ^{\ell +1}r^{\frac{1}{2}}%
\sqrt{2\lambda \Gamma \left( \ell +\frac{1}{2}\right) }e^{\lambda
^{-2}E}\left( 2E\right) ^{-\frac{1}{2}\left( \ell +\frac{1}{2}\right)
}J_{\ell +\frac{1}{2}}\left( \sqrt{2r^{2}E}\right).   \tag{4.12}
\end{equation}%
We also need to specify the kernel function $\mathcal{S}$ defined in $\left(
2.30\right) :$ 
\begin{equation}
\mathcal{S}\left( c_{k},y\right) :=\sum_{n=0}^{k}Q_{n}\left( c_{k}\right)
P_{n}\left( y\right) =\sum_{n=0}^{k}\prod\limits_{j=0}^{n-1}\left( \frac{%
c_{k}-c_{j}}{d_{j+1}}\right) P_{n}\left( y\right) .  \tag{4.13}
\end{equation}%
Therefore,%
\begin{equation}
\langle r\left\vert c_{k},\gamma \right\rangle =\left( \mathcal{N}\left(
c_{k}\right) \right) ^{-\frac{1}{2}}\sum_{n=0}^{k}\prod\limits_{j=0}^{n-1}%
\left( \frac{c_{k}-c_{j}}{d_{j+1}}\right) \int\limits_{0}^{+\infty }\mathcal{%
K}\left( r,y\right) P_{n}\left( y\right) \omega _{\lambda ,\ell }\left(
y\right) e^{-i\gamma y}dy.  \tag{4.14}
\end{equation}%
The integral in $\left(4.14\right) $ reads%
\begin{equation*}
\int\limits_{0}^{+\infty }\mathcal{K}\left( r,y\right) P_{n}\left( y\right)
\varrho \left( y\right) e^{-i\gamma y}dy=\frac{\left( -1\right) ^{n}\lambda
^{-\ell -\frac{3}{2}}}{\left( \ell +\frac{1}{2}\right) }\sqrt{\frac{2n!}{%
\Gamma \left( n+\ell +\frac{3}{2}\right) }}r^{\frac{1}{2}}2^{\frac{1}{2}%
\left( \ell +\frac{1}{2}\right) }
\end{equation*}

\begin{equation}
\times \int\limits_{0}^{+\infty }e^{-\frac{1}{\lambda ^{2}}y}y^{\frac{1}{2}%
\left( \ell +\frac{1}{2}\right) }J_{\ell +\frac{1}{2}}\left( \sqrt{2r^{2}y}%
\right) L_{n}^{\left( \ell +\frac{1}{2}\right) }\left( \frac{2}{\lambda ^{2}}%
y\right) e^{-i\gamma y}dy.  \tag{4.15}
\end{equation}%
For $k=0$ formula $\left( 4.14\right) $ reduces to 
\begin{equation}
\langle r\left\vert c_{0},\gamma \right\rangle =\left( \mathcal{N}\left(
c_{0}\right) \right) ^{-\frac{1}{2}}\int\limits_{0}^{+\infty }\mathcal{K}%
\left( r,y\right) \omega _{\lambda ,\ell }\left( y\right) e^{-i\gamma y}dy 
\tag{4.16}
\end{equation}%
\begin{equation*}
=\left( \mathcal{N}\left( c_{0}\right) \right) ^{-\frac{1}{2}}\frac{1}{2}%
\lambda ^{\ell +1}r^{\frac{1}{2}}\sqrt{2\lambda \Gamma \left( \ell +\frac{1}{%
2}\right) }\frac{2}{\lambda ^{2}}\frac{1}{\Gamma \left( \ell +\frac{3}{2}%
\right) }2^{-\frac{1}{2}\left( \ell +\frac{1}{2}\right) }\left( \frac{2}{%
\lambda ^{2}}\right) ^{\frac{1}{2}+\ell }
\end{equation*}%
\begin{equation}
\times \int\limits_{0}^{+\infty }y^{\frac{1}{2}\left( \ell +\frac{1}{2}%
\right) }J_{\ell +\frac{1}{2}}\left( \sqrt{2r^{2}y}\right) e^{-\frac{1}{%
\lambda ^{2}}y}e^{-i\gamma y}dy.  \tag{4.17}
\end{equation}%
By the variable change $y=x^{2},$ the last integral becomes 
\begin{equation}
2\int\limits_{0}^{+\infty }x^{\left( \ell +\frac{1}{2}\right) +1}J_{\ell +%
\frac{1}{2}}\left( \sqrt{2r^{2}}x\right) e^{-(\frac{1}{\lambda ^{2}}+i\gamma
)x^{2}}dx.  \tag{4.18}
\end{equation}%
Next, by using the identity (\cite{GR}, p.706) 
\begin{equation}
\int\limits_{0}^{+\infty }x^{\nu +1}e^{-\alpha x^{2}}J_{\nu }\left( \beta
x\right) dx=\frac{\beta ^{\nu }}{\left( 2\alpha \right) ^{\nu +1}}\exp
\left( -\frac{\beta ^{2}}{4\alpha }\right) ,\text{ \ \ }\func{Re}\alpha >0,%
\func{Re}\nu >-1,  \tag{4.19}
\end{equation}%
for $\beta =\sqrt{2r^{2}},\nu =\left( \ell +\frac{1}{2}\right) $ and $\alpha
=(\frac{1}{\lambda ^{2}}+i\gamma )$, we arrive at the expression

\begin{equation}
\langle r\left\vert \lambda ,\gamma \right\rangle =\sqrt{2}\left( \Gamma
\left( \ell +\frac{3}{2}\right) \right) ^{-\frac{1}{2}}\left( \frac{1}{%
\lambda }\right) ^{\left( \ell +\frac{3}{2}\right) }\frac{r^{\ell +1}}{%
\left( \frac{1}{\lambda ^{2}}+i\gamma \right) ^{\ell +\frac{3}{2}}}\exp
\left( -\frac{r^{2}}{2(\frac{1}{\lambda ^{2}}+i\gamma )}\right) .  \tag{4.20}
\end{equation}%
Note that with respect to the basis $\left( 4.2\right) $ one can observe
that the coefficient $c_{0}$ in $\left( 4.16\right) $ coincides with the
labeling parameter $z$ according to calculations that start by the formula $%
\left( 2.16\right) .$ So the above equation $\left( 4.20\right) $ represents
in fact the wave function in $r-$coordinate of a coherent state with the
given $z$ provided we choose the value $\lambda =\frac{2}{\sqrt{2l+3}}z.$

Now, for a Bayesian decomposition of the weight function purpose, we first
observe that $\omega _{\lambda ,\ell }\left( E\right) $ as given by $\left(
4.7\right) $ is a Gamma distribution $\mathcal{G}\left( \ell +\frac{3}{2},%
\frac{2}{\lambda ^{2}}\right) .$ It is also well known that for the Poisson
model $X\sim \mathcal{P}\left( \kappa \right) $ with $\kappa >0$ given by
the probability distribution \textit{\ }%
\begin{equation}
\Pr \left( X=j\right) =\frac{\kappa ^{j}}{j!}e^{-\kappa },\text{ }%
j=0,1,2,...,  \tag{4.21}
\end{equation}%
\textit{\ }if the prior distribution on the parameter $\kappa $ is a Gamma
distribution $\mathcal{G}\left( \alpha ,\beta \right) $\ then the posterior
distribution is also a Gamma distribution\textit{\ }$\mathcal{G}\left(
\alpha +j,\beta +1\right) .$ Thus, in terms of our notations, 
\begin{equation}
\Pr \left( X=\ell \right) =\frac{E^{\ell }}{\ell !}e^{-E}\equiv p_{E}\left(
l\right) ,\text{ }l=0,1,2,...,\text{ }E>0\text{\ }  \tag{4.22}
\end{equation}%
where $X\sim \mathcal{P}\left( E\right) ,$ is a convenient statistical
model. This also indicates that the angular momentum\ integer number $l$ may
in fact play the role of an observed data of a discrete random variable \ $%
X\sim \mathcal{P}\left( E\right) $ with the energy $E>0$ as its parameter.
Therefore, we now fixe $l$ and proceed to reverse $X$ by fixing\textit{\ "%
\`{a} priori" }a law $\pi _{\lambda }\left( E\right) $ that $E$ is supposed
to follow. The prior law on the prameter can be obtained just by writting
our weight function $\omega _{\lambda ,\ell }\left( E\right) \equiv \mathcal{%
G}\left( \ell +\frac{3}{2},\frac{2}{\lambda ^{2}}\right) $ as a gamma
distribution $\mathcal{G}\left( \alpha +\ell ,\beta +1\right) .$This gives us%
\begin{equation}
\pi _{\lambda }\left( E\right) :=\mathcal{G}\left( \frac{3}{2},\frac{2}{%
\lambda ^{2}}-1\right)   \tag{4.23}
\end{equation}%
and therefor we can rewrite the weight function as a posterior distribution
as 
\begin{equation}
\frac{\left[ \pi _{\lambda }\left( E\right) \right] \left[ p_{E}\left( \ell
\right) \right] }{\int \left[ \pi _{\lambda }\left( y\right) \right] \left[
p_{y}\left( \ell \right) \right] dy}=\omega _{\lambda ,l}\left( E\right), 
\tag{4.24}
\end{equation}%
which, after simplification, reduces to%
\begin{equation}
\frac{\left[ e^{-\frac{2}{\lambda ^{2}}E}\right] \left[ E^{\frac{1}{2}+\ell }%
\right] }{\int\limits_{0}^{+\infty }y^{\frac{1}{2}+\ell }e^{-\frac{2}{%
\lambda ^{2}}y}dy}=\omega _{\lambda ,\ell }\left( E\right).   \tag{4.25}
\end{equation}%
In other words, 
\begin{equation}
\omega _{\lambda ,l}\left( E\right) \propto \left( \tau \left( \lambda
\right) \right) ^{2E}q\left( E\right)   \tag{4.26}
\end{equation}%
where 
\begin{equation}
\tau \left( \lambda \right) =e^{-\frac{1}{\lambda ^{2}}}  \tag{4.27}
\end{equation}%
and 
\begin{equation}
q\left( E\right) =E^{\frac{1}{2}+l}.  \tag{4.27}
\end{equation}%
Now, in order to recover the constructed CS in $\left( 4.20\right) $ by the
Gazeau-Klauder formalism let us recall that the operator $H_{\ell }$ acts on
the Hilbert space $L^{2}\left( \mathbb{R}_{+},dr\right) $ and admits a
continous spectrum $E\in \left[ 0,+\infty \right) $. The Schr\"{o}dinger
equation $H_{\ell }\varphi =E\varphi $ has a regular solution given by 
\begin{equation}
\hat{\jmath}_{\ell }\left( kr\right) =\sqrt{kr}J_{\ell +\frac{1}{2}}\left(
kr\right) ,\text{ \ }  \tag{4.28}
\end{equation}%
where $J_{\nu }$ denotes the Bessel function of the first kind and of order $%
\nu $ (\cite{GR}, p.910) and $k=\sqrt{2E}$. The function $\hat{\jmath}_{\ell
}$ is regular for $r\rightarrow 0$ for $\ell >0.$ Therefore, eigenstates are
those given by 
\begin{equation}
\langle r\left\vert E\right\rangle =\sqrt{kr}J_{\ell +\frac{1}{2}}\left(
kr\right) .  \tag{4.29}
\end{equation}%
From the above Bayesian decomposition of the weight function $\omega _{\ell
,\lambda }\left( E\right) $ we choose the factorial function to be defined
by 
\begin{equation}
f_{\ell }\left( E\right) :=\frac{1}{q\left( E\right) }=\left( 2E\right)
^{-\left( \frac{1}{2}+\ell \right) }.  \tag{4.30}
\end{equation}%
Therefore, the corresponding Steiljes moment problem%
\begin{equation}
f_{l}(E)=\int\limits_{0}^{+\infty }s^{2E}\sigma (s)ds  \tag{4.31}
\end{equation}%
can be solved by the weight function 
\begin{equation}
\sigma _{\ell }\left( s\right) =\frac{1}{\Gamma \left( \frac{1}{2}+\ell
\right) }\frac{1}{s}\left( Log\frac{1}{s}\right) ^{-\frac{1}{2}+\ell },\text{
}s<1  \tag{4.32}
\end{equation}%
and $\sigma _{\ell }\left( s\right) =0$ for $s\geq 1,$ by making appeal to
the Mellin transform (\cite{Bat}, p.343) : 
\begin{equation}
\int\limits_{0}^{+\infty }\phi _{\alpha ,\nu }\left( x\right)
x^{p-1}dx=\Gamma \left( \nu \right) \left( p+1\right) ^{-\nu },\func{Re}\nu
>0,\func{Re}p>-\func{Re}\alpha ,  \tag{4.33}
\end{equation}%
where $\phi _{\alpha ,\nu }\left( x\right) =x^{\alpha }\left( -Logx\right)
^{\nu -1}$, $0<x<1$ and $\phi _{\alpha ,\nu }\left( x\right) =0$, $x\in %
\left[ 1,+\infty \right) ,$ for $p=2E+1,\nu =\frac{1}{2}+\ell $ and $\alpha
=-1$. Therefore, the normalization factor $\left( 2.3\right) $, here, 
\begin{equation}
\mathcal{N}_{\ell }\left( s\right) =\frac{1}{2}\Gamma \left( \frac{3}{2}%
+\ell \right) \left( Log\frac{1}{s}\right) ^{-\left( \frac{3}{2}+\ell
\right) }.  \tag{4.34}
\end{equation}%
With these ingredients, the CS $\left( 3.7\right) $ take the
form%
\begin{equation}
\left\vert s,\gamma \right\rangle =\left( \mathcal{N}_{\ell }\left( s\right)
\right) ^{-\frac{1}{2}}\int\limits_{0}^{+\infty }dE\frac{s^{E}e^{-i\gamma E}%
}{\sqrt{\left( 2E\right) ^{-\left( \frac{1}{2}+\ell \right) }}}\left\vert
E\right\rangle .  \tag{4.35}
\end{equation}%
Now, from the above Bayesian decomposition of $\omega _{\ell ,\lambda
}\left( E\right) $ we choose the following reparametrization for the
labeling parameter $s$ according to $\left( 4.27\right) $ as 
\begin{equation}
s=\tau \left( \lambda \right) =\exp \left( -\frac{1}{\lambda ^{2}}\right) ,%
\text{ \ }0\leq s<1,\text{ \ }\lambda \in \mathbb{R},  \tag{4.36}
\end{equation}%
then $\left( 4.18\right) $ takes the form 
\begin{equation}
\left\vert \lambda ,\gamma \right\rangle =\left( \mathcal{N}_{l}\left(
s\right) \right) ^{-\frac{1}{2}}\int\limits_{0}^{+\infty }\frac{e^{-(\frac{1%
}{\lambda ^{2}}+i\gamma )E}}{\sqrt{\left( 2E\right) ^{-\left( \frac{1}{2}%
+\ell \right) }}}\left\vert E\right\rangle dE.  \tag{4.37}
\end{equation}%
Next, making use of $\left( 4.29\right) $, we obtain successively%
\begin{equation}
\langle r\left\vert \lambda ,\gamma \right\rangle =\left( \mathcal{N}%
_{l}\left( s\right) \right) ^{-\frac{1}{2}}\int\limits_{0}^{+\infty }\left(
2E\right) ^{\frac{1}{2}\left( \frac{1}{2}+l\right) }e^{-(\frac{1}{\lambda
^{2}}+i\gamma )E}\left[ \sqrt{r}J_{l+\frac{1}{2}}\left( kr\right) \right] dE
\tag{4.38}
\end{equation}%
\begin{equation}
=\sqrt{r}\left( \mathcal{N}_{l}\left( s\right) \right) ^{-\frac{1}{2}%
}\int\limits_{0}^{+\infty }\sqrt{2E}^{\left( \frac{1}{2}+l\right) }e^{-\frac{%
1}{2}(\frac{1}{\lambda ^{2}}+i\gamma )\left( \sqrt{2E}\right) ^{2}}J_{l+%
\frac{1}{2}}\left( \sqrt{2E}r\right) dE  \tag{4.39}
\end{equation}%
\begin{equation}
=\sqrt{r}\left( \mathcal{N}_{l}\left( s\right) \right) ^{-\frac{1}{2}%
}\int\limits_{0}^{+\infty }x^{(\frac{1}{2}+l)+1}e^{-\frac{1}{2}(\frac{1}{%
\lambda ^{2}}+i\gamma )x^{2}}J_{l+\frac{1}{2}}\left( xr\right) dx.  \tag{4.40}
\end{equation}%
By applying the formula (\cite{GR}, p.706):%
\begin{equation}
\int\limits_{0}^{+\infty }x^{\nu +1}e^{-\alpha x^{2}}J_{\nu }\left( \beta
x\right) dx=\frac{\beta ^{\nu }}{\left( 2\alpha \right) ^{\nu +1}}\exp
\left( -\frac{\beta ^{2}}{4\alpha }\right) ,\text{ \ \ }\func{Re}\alpha >0,%
\func{Re}\nu >-1,  \tag{4.41}
\end{equation}%
for parameters $\nu =l+\frac{1}{2},$   $\alpha =\frac{1}{2}(\frac{1}{%
\lambda ^{2}}+i\gamma )$ and\ $\beta =r$, Eq. $\left( 4.40\right) $ reads 
\begin{equation}
\langle r\left\vert \lambda ,\gamma \right\rangle =\left( \mathcal{N}%
_{l}\left( s\right) \right) ^{-\frac{1}{2}}\frac{r^{l+1}}{\left( \frac{1}{%
\lambda ^{2}}+i\gamma \right) ^{l+\frac{1}{2}+1}}\exp \left( -\frac{r^{2}}{2(%
\frac{1}{\lambda ^{2}}+i\gamma )}\right) .  \tag{4.42}
\end{equation}%
Finally, we replace $\mathcal{N}_{l}\left( s\right) $ by is expression $%
\left( 4.34\right) $ to arrive at the expression

\begin{equation}
\langle r\left\vert \lambda ,\gamma \right\rangle =\sqrt{2}\left( \Gamma
\left( l+\frac{3}{2}\right) \right) ^{-\frac{1}{2}}\left( \frac{1}{\lambda }%
\right) ^{\left( l+\frac{3}{2}\right) }\frac{r^{l+1}}{\left( \frac{1}{%
\lambda ^{2}}+i\gamma \right) ^{l+\frac{3}{2}}}\exp \left( -\frac{r^{2}}{2(%
\frac{1}{\lambda ^{2}}+i\gamma )}\right).   \tag{4.43}
\end{equation}%
The above expression of coherent states $\left( 4.42\right) $ is a major
result. It has the following properties. For $\gamma =0,$ the corresponding
expression of CS reduces to 
\begin{equation}
\langle r\left\vert \lambda ,0\right\rangle =\sqrt{\frac{2\lambda }{\Gamma
\left( \ell +\frac{3}{2}\right) }}(\lambda r)^{l+1}\exp \left( -\frac{%
\lambda ^{2}}{2}r^{2}\right).   \tag{4.44}
\end{equation}%
Recall that for the basis vectors $\phi _{n}^{(\ell ,\lambda )}\left(
r\right) $ in $\left( 4.2\right) $ we have for $n=0$%
\begin{equation}
\phi _{0}^{(\ell ,\lambda )}\left( r\right) :=\sqrt{\frac{2\lambda }{\Gamma
\left( \ell +\frac{3}{2}\right) }}\left( \lambda r\right) ^{\ell +1}\exp
\left( -\frac{\lambda ^{2}}{2}r^{2}\right) .  \tag{4.45}
\end{equation}%
So we may rewrite $\left( 4.44\right) $ as $\langle r\left\vert \lambda
,0\right\rangle =\phi _{0}^{(\ell ,\lambda )}\left( r\right) $ as expected.
The combined energy exponential in the integral in $\left( 4.37\right) $ is
now 
\begin{equation}
e^{-\frac{1}{\beta ^{2}}E},\text{ \ \ \ }\frac{1}{\beta ^{2}}=\frac{1}{%
\lambda ^{2}}+i\gamma.   \tag{4.46}
\end{equation}%
We therefore have the result 
\begin{equation}
\langle r\left\vert \lambda ,\gamma \right\rangle =\left( \frac{\beta }{%
\lambda }\right) ^{\ell +\frac{3}{2}}\phi _{0}^{(\ell ,\beta )}\left(
r\right) .  \tag{4.47}
\end{equation}%
Explicitly, we have the density function\textbf{\ }%
\begin{equation}
\rho \left( r;\lambda ,\gamma \right) :=\left\vert \langle r\left\vert
\lambda ,\gamma \right\rangle \right\vert ^{2}=\frac{2\lambda ^{2l+3}}{%
\Gamma \left( l+\frac{3}{2}\right) (1+\gamma ^{2}\lambda ^{4})^{l+3/2}}%
\;r^{2l+2}e^{\frac{-r^{2}\lambda ^{2}}{1+\lambda ^{4}\gamma ^{2}}}.  \tag{4.48}
\end{equation}%
In figure 1, we show the behavior of the function $r\mapsto \rho \left(
r;\lambda ,\gamma \right) $ for several discrete values of $\gamma .$ 
\begin{figure}[tbp]
\centering
\includegraphics[scale=0.3]{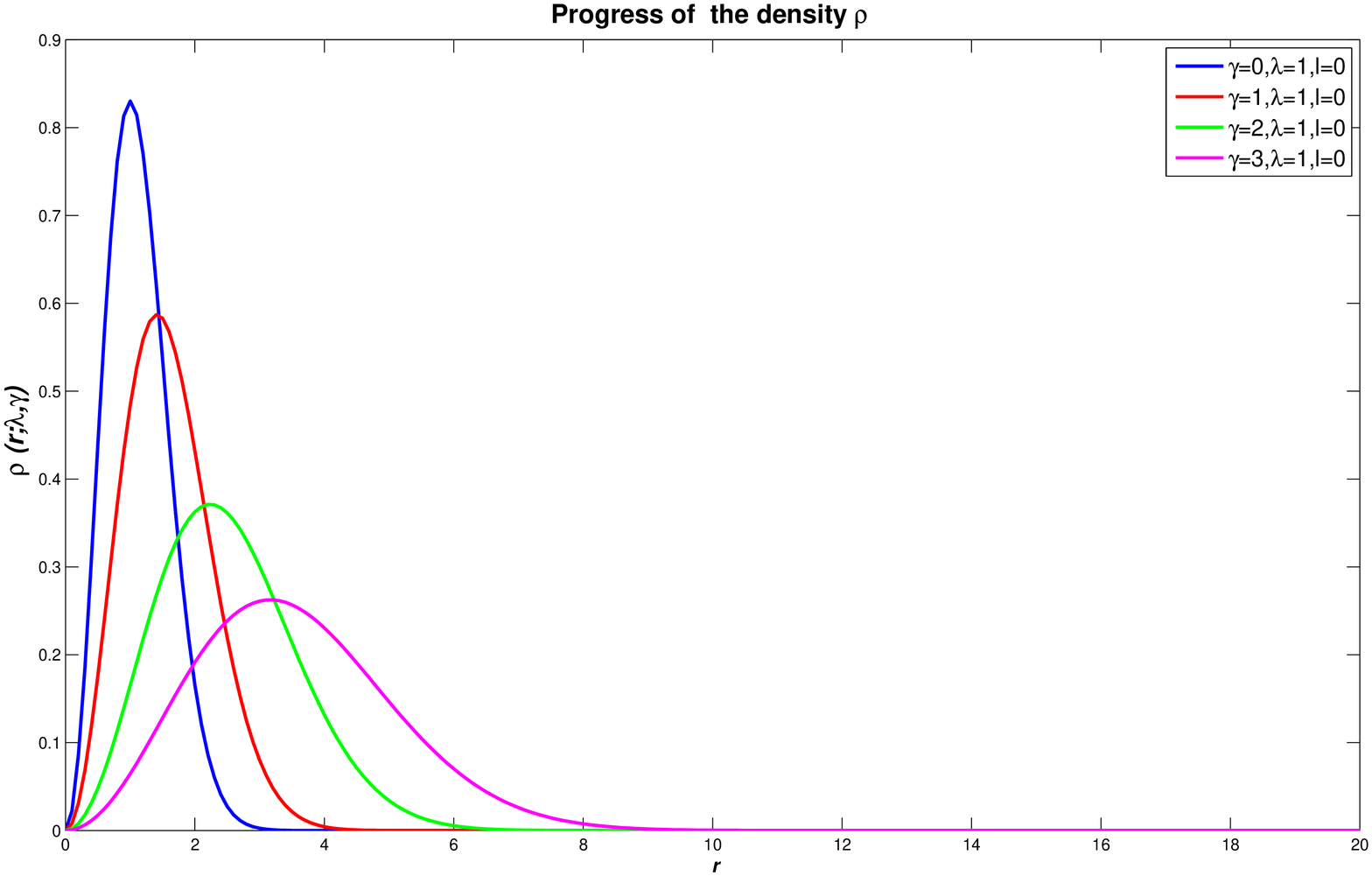}
\end{figure}
\begin{figure}[tbp]
\centering
\includegraphics[scale=0.3]{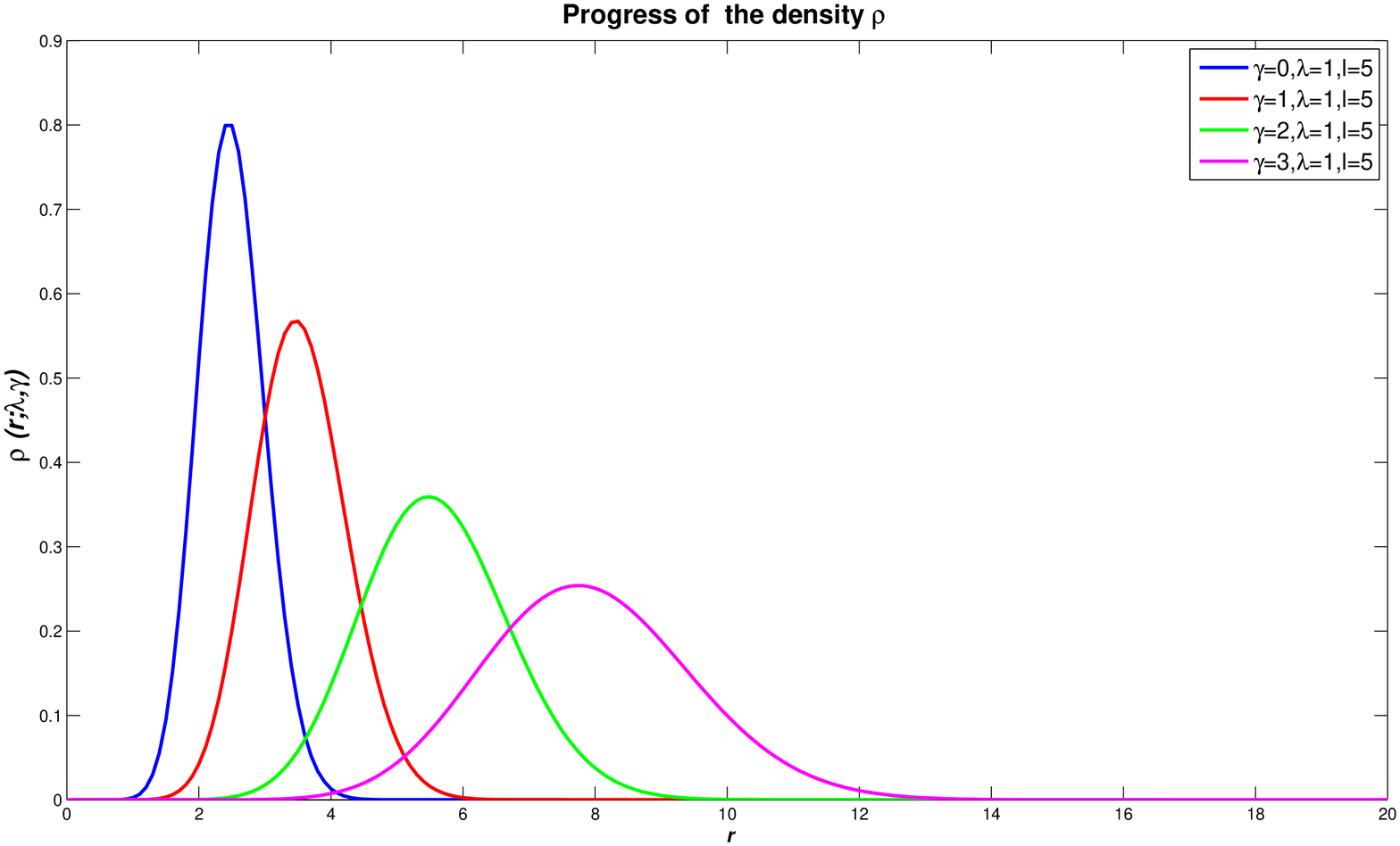}
\end{figure}
\begin{figure}[tbp]
\centering
\includegraphics[scale=0.3]{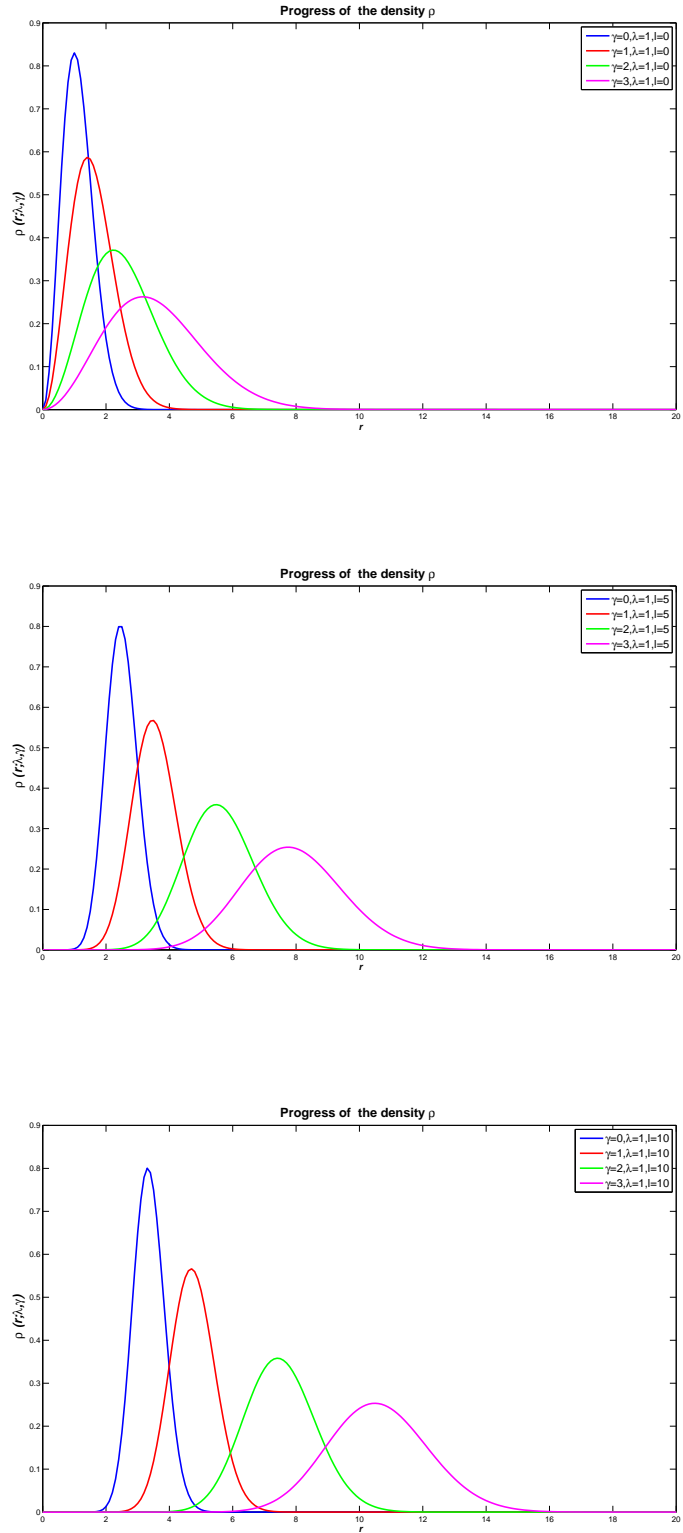}
\caption{$r\mapsto \protect\rho \left( r;\protect\lambda ,\protect\gamma %
\right) $ for several discrete values of $\protect\gamma .$ }
\label{fig:example}
\end{figure}
We now can calculate the average position 
\begin{equation}
\overline{r}\left( \gamma \right) :=\int\limits_{0}^{+\infty }r\rho \left(
r;\lambda ,\gamma \right) dr=\frac{2\lambda ^{2l+3}}{\Gamma \left( l+\frac{3%
}{2}\right) (1+\gamma ^{2}\lambda ^{4})^{l+3/2}}\int\limits_{0}^{+\infty }e^{%
\frac{r^{2}}{\lambda ^{2}\gamma ^{2}-\lambda ^{-2}}}r^{2l+3}dr.  \tag{4.49}
\end{equation}%
Applying the integral (\cite{GR}, p.337): 
\begin{equation}
\int\limits_{0}^{+\infty }x^{m}e^{-\beta x^{n}}dx=\frac{1}{n\beta ^{\frac{m+1%
}{n}}}\Gamma (\frac{m+1}{n}),\text{ \ }\func{Re}m>0,\,\func{Re}\,n>0,\,\func{%
Re}\beta >0  \tag{4.50}
\end{equation}\medskip
for $m=2l+3$, $n=2$ and $\beta =\frac{\lambda ^{2}}{1+\lambda ^{4}\gamma ^{2}%
}$, Eq.$\left( 4.49\right) $ takes the form 
\begin{equation}
\overline{r}\left( \gamma \right) =C\frac{(l+1)!}{\Gamma \left( l+3/2\right) 
}\left( \frac{1}{\lambda ^{2}}+\lambda ^{2}\gamma ^{2}\right) ^{1/2}. 
\tag{4.51}
\end{equation}%
Note that when $l=0$, the average position reduces to $\frac{2}{\sqrt{\pi }}%
\left( \frac{1}{\lambda ^{2}}+\lambda ^{2}\gamma ^{2}\right) ^{1/2}$%
.  On other hand the velocity (with respect to $\gamma )$ is%
\begin{equation*}
\upsilon \left( \gamma \right) :=\partial _{\gamma }\left( \overline{r}%
\left( \gamma \right) \right) =\frac{(l+1)!}{\Gamma \left( l+3/2\right) }%
\frac{\lambda ^{2}\gamma }{\sqrt{\frac{1}{\lambda ^{2}}+\lambda ^{2}\gamma
^{2}}}.
\end{equation*}%
Figure 2 shows how quickly this velocity reaches its asymptotic value
as $\gamma $ goes to infinity:%
\begin{equation*}
\lim_{\gamma \rightarrow +\infty }\upsilon \left( \gamma \right) =\lambda 
\frac{\Gamma (l+2)}{\Gamma \left( l+3/2\right) }.
\end{equation*}
\begin{figure}[tbp]
\centering
\includegraphics[scale=0.3]{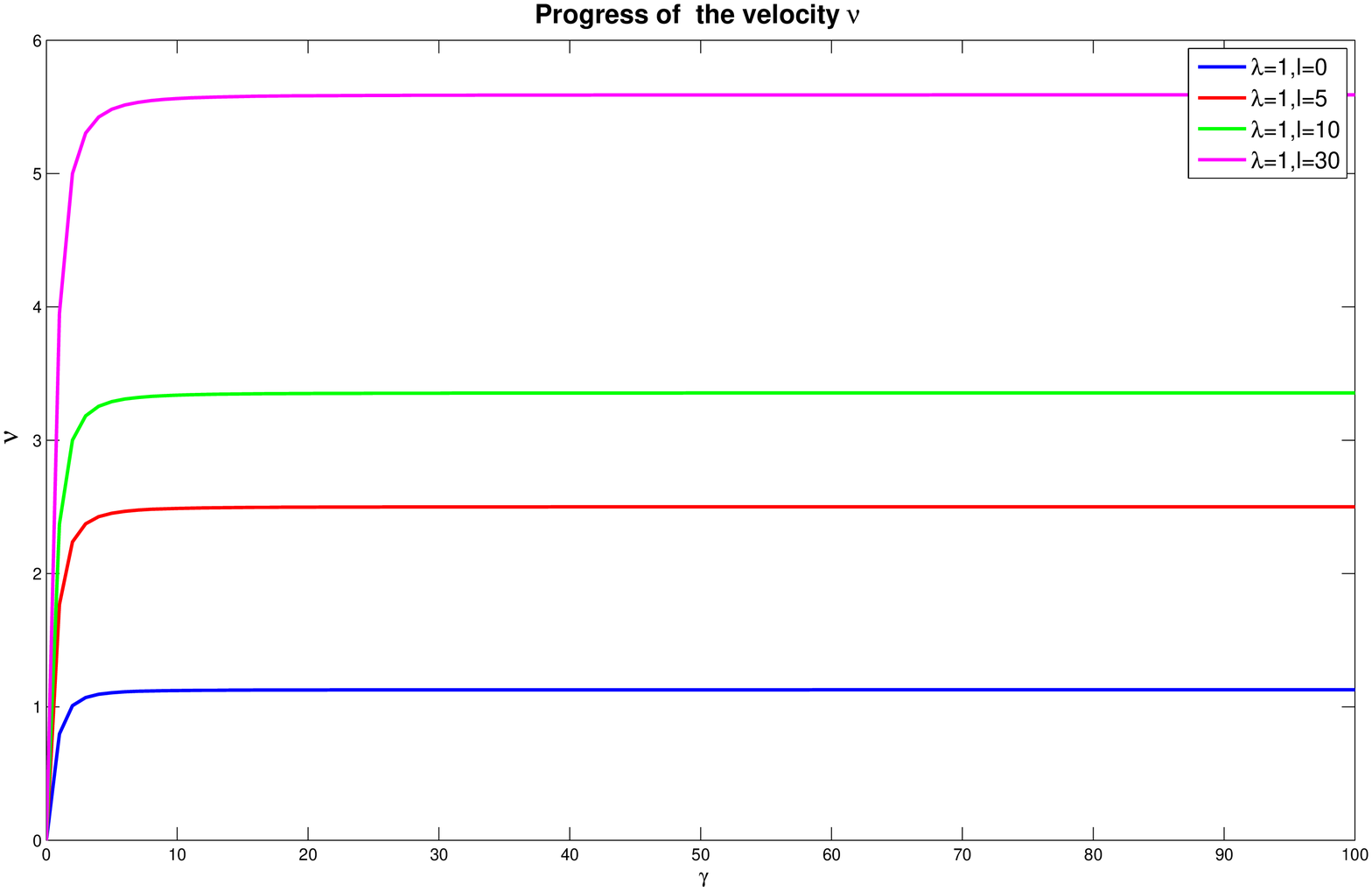}
\caption{{}}
\end{figure}
\section{\textbf{Concluding remarks}}
We have constructed a set of CS obeying a Glauber-type condition for a
Hamiltonian with continuous spectrum by using a tridiagonal method involving
orthogonal polynomials. The basic quantities in our procedure are the
parameters $\left( c_{n},d_{n}\right) $ which are related to the matrix
elements $\left( a_{n},b_{n}\right) $ of the tridiagonal Hamiltonian by $%
\left( 2.13\right) $ and $\left( 2.14\right) .$ More specifically, these CS
are labeled by the sequence $z=c_{n}.$ But the general form $\left(
2.16\right) $ is still to be exploited. Connecting these states with the
Gazeau-Klauder CS was not straightforward and bridge the gap between the two
approaches requires the idea of a Bayesian decomposition for the weight
function in the orthogonality measure of polynomials arising from the
tridiagonal method. As an example, we have the $\ell $-wave free particle
for which the statistical model given by the Poisson probability
distribution $\Pr \left( X=l\right) =e^{-E}E^{l}/l!,$ $l=0,1,2,...,$ \ $%
X\sim \mathcal{P}\left( E\right) $, has played a central role in writing
down the convenient Bayesian decomposition for the corresponding weight
function. Therefore, there should be an explanation for the appearance of
the Poisson distribution having the energy $E>0$\ as a parameter and the
set of all angular momentum numbers $l$ as its observed data in the physics
of this system.

\end{document}